# Possible martensitic transformation in $Pd_2MnTi$ and $Pt_2MnTi$: First-principles investigation

L. Feng

*Department of Physics, Taiyuan University of Technology, 030024, Taiyuan, Republic of China*

**Abstract:** The martensitic transformation in all-*d*-metal Heusler alloys $Pd_2MnTi$ and $Pt_2MnTi$ have been investigated based on first-principles investigations. The calculated results indicate that the martenstic transformation have great possibility to occur in both $Pd_2MnTi$ and $Pt_2MnTi$. The energy differences between the cubic and tetragonal phases are 215.12 meV and 329.45 meV for $Pd_2MnTi$ and $Pt_2MnTi$, respectively. The analysis of the electronic structure of cubic and tetragonal phases also support this conclusion. The magnetic properties are also investigated for the two compounds.

**Introduction:**

Magnetic shape memory alloys have many applications, such as magnetic-field-induced martensitic transformation, magneto-induced strain, magnetocaloric effect, exchange bias and so on. Most magnetic shape memory alloys are Heusler alloys. And it has been found that, in these magnetic shape memory alloys, *p-d* covalent hybridization plays an important role in atomic ordering, lattice structure stability and long-range magnetic ordering. Recently, all-*d*-metal Heusler alloys have drawn great interest in the field of the magnetic material [1-5]. Generally, the chemical formula of Heusler alloys is $X_2YZ$, in which X and Y are transition metals, and Z is main group element. And when the Z is also transition metal, all-*d*-metal Heusler is obtained. It has already known that the *p-d* interaction have great influence on the electronic structure and the stability of the Heusler alloy. What will happen when the *d-d* interaction replace the *p-d* interaction. The work by E. K. Liu *et. al* has indicated that the martensitic transformation can still occur in Ni-Mn-Ti Heusler alloy [1]. In addition, the physical effects related to magnetism and martensitic transformation are also observed, such as magnetocaloric effect [2, 3], magnetoresistance effect [4], and so on. Thus, this discovery opens up a new field of magnetic materials. Since Pd and Pt belong to the same group as Ni in the chemical periodic table, it is worth studying whether the martensitic transformation can occur in Pd-Mn-Ti and Pt-Mn-Ti. In this work, we have investigated the possibility of occurrence for the martensitic transformation in $Pd_2MnTi$ and $Pt_2MnTi$ Heusler alloys using first-principles calculations. The electronic structure and the magnetic property have also given out. Based on these results, these two compounds might be synthesized and studied experimentally.

2. Computational details

The calculations are performed by Cambridge Serial Total Energy Package (CASTEP) code [6]. The Perdew-Wang generalized gradient approximation [7] is used to describe the exchange correlation energy. The ultrasoft pseudopotentials [8] are used to describe the interactions between ion cores and valence electrons. For the pseudopotentials used, the electronic configurations with core level correction are $Pd(4d^{10})$, $Pt(5d^96s^1)$, $Mn(3d^54s^2)$ and $Ti(3d^24s^2)$, respectively. The cut-off energy of the plane wave basis set is 500 eV for all of the cases. The scheme for generating k-points is Monkhorst-Pack method, and k-points are set as 20×20×20 and 18×18×10 in the irreducible Brillouin zone of cubic and the tetragonal phases, respectively. The convergence criterion for the calculations is selected as the total energy difference within $10^{-6}$ eV/atom.

3. Results and discussions.

3.1 Crystal structure.

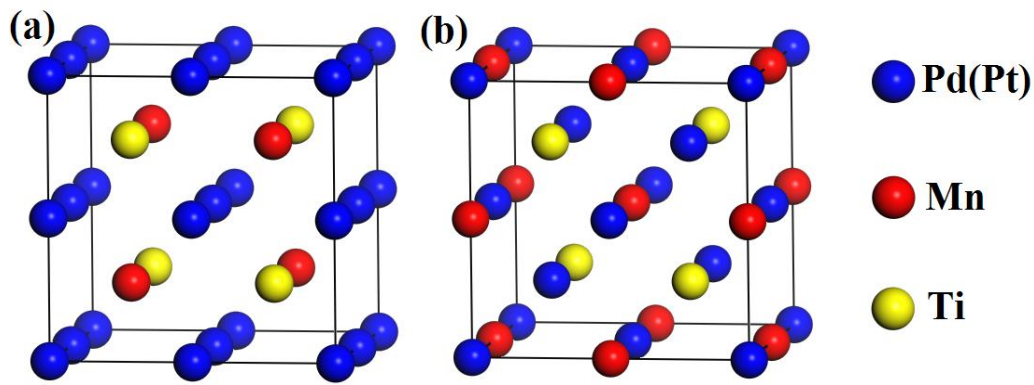

Fig. 1 Atomic configuration of Pd$_2$MnTi and Pt$_2$MnTi with L2$_1$ and Hg$_2$CuTi structure

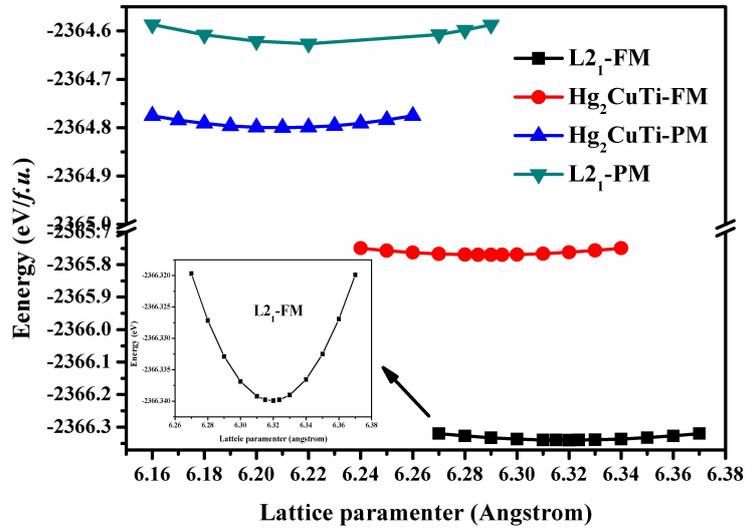

Fig. 2 Total energy of Pt$_2$MnTi as a function of lattice parameter

Both L2$_1$ structure and Hg$_2$CuTi structure are considered to determine the ground state, as shown in Fig. 1. In L2$_1$ structure, Pt(Pd) atoms occupy A (0, 0, 0) and C (1/2, 1/2, 1/2) sites, and Ti and Mn atoms occupy B (1/4, 1/4, 1/4) and D (3/4, 3/4, 3/4) sites, respectively. For Hg$_2$CuTi structure, however, Pt(Pd) atoms occupy A (0, 0, 0) and B (1/4, 1/4, 1/4) sites, while Mn and Ti atoms occupy C (1/2, 1/2, 1/2) and D (3/4, 3/4, 3/4) sites. To obtain the ground state of the austenitic phase of Pd$_2$MnTi and Pt$_2$MnTi, both the ferromagnetic (FM) and antiferromagnetic (AFM) of L2$_1$ structure and Hg$_2$CuTi structure are calculated. As a result, for Pd$_2$MnTi, we obtain the AFM results even though we set the initial magnetic structure to be parallel (FM). while for Pt$_2$MnTi, we obtain the FM results even though we set the initial magnetic structure to be antiparallel (AFM). For both the L2$_1$ structure and the Hg$_2$CuTi structure, the magnetic state is more stable than the paramagnetic state (PM) (as shown in Fig. 2). It can also be found that, for Pt$_2$MnTi, the total energy of the ferromagnetic state in the L2$_1$ structure is the lowest, which indicates that the stable structure of the austenitic phase of Pt$_2$MnTi is the ferromagnetic L2$_1$ structure. And the stable structure of the austenitic phase of Pd$_2$MnTi is the antiferromagnetic L2$_1$ structure. And the lattice parameters for Pd$_2$MnTi and Pt$_2$MnTi are 6.30 Å and 6.32 Å, respectively.

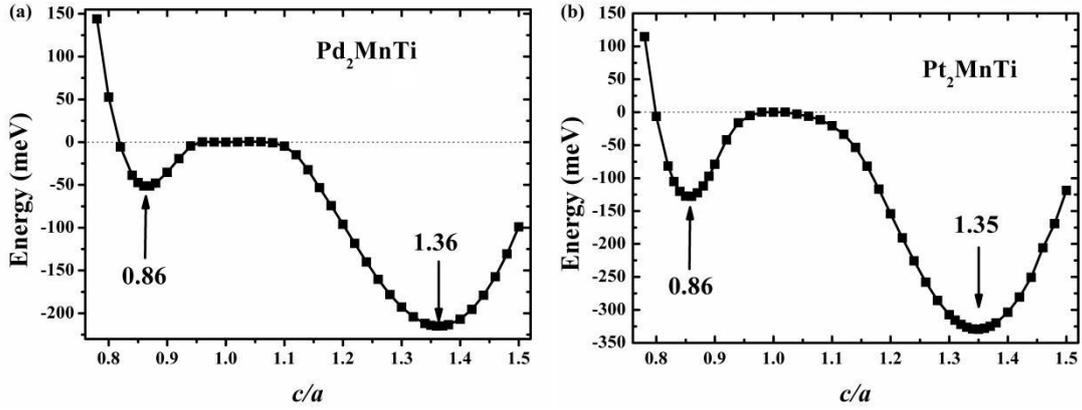

Fig. 3 (a) Variation of the total energy of Pd$_2$MnTi with *c/a*; (b) Variation of the total energy of Pt$_2$MnTi with *c/a*

In order to investigate the possibility of occurrence for the martensitic transformation, the volume-conserving tetragonal distortion has been conducted for Pd$_2$MnTi and Pt$_2$MnTi. It means that we keep the cell volume constant and change the *c/a* ratio of the tetragonal phase. And the total energy of all the tetragonal phases with different *c/a* ratio. Here, the c and a are the lattice parameter of the distorted tetragonal phase. The variation of the total energy of Pd$_2$MnTi with *c/a* ratio has shown in Fig. 3(a). It can be found that there are two minimums for the energy curve of Pd$_2$MnTi. And one minimum locates at *c/a*=0.86 and the other one locates at *c/a*=1.36. In addition, the energy of the tetragonal phase with *c/a*=1.36 is the lowest. The energy differences between the two minimums and the energy of the cubic phase are 50 meV/*f.u.* and 200 meV/*f.u.* Based on the a large amount of simulation and experimental results, a empirical rule has been obtained to predict the martensitic transformation of the Heusler alloy: the distortion degree is in the appropriate range of 1.2-1.4 and the difference between the cubic phase and the tetragonal phase is large enough to overcome the resistance of the structure distortion. For example, the calculated results for Ni$_2$MnGa indicate that the *c/a* ratio of the predicted tetragonal phase is 1.26. And the energy difference between the cubic and tetragonal phase is 32 meV/*f.u.* [9]. Obviously, the situation in Pd$_2$MnTi is conforms to the empirical rule. And the martensitic transformation has great possibility to occur in Pd$_2$MnTi. Similar analysis also applies to Pt$_2$MnTi. The variation of the total energy of Pt$_2$MnTi with *c/a* ratio has shown in Fig. 3(b). It can be found that there are also two minimums for the energy curve of Pt$_2$MnTi. And one minimum locates at *c/a*=0.86 and the other one locates at *c/a*=1.35. In addition, the energy of the tetragonal phase with *c/a*=1.35 is the lowest. The energy differences between the two minimums and the energy of the cubic phase are 125 meV/*f.u.* and 325 meV/*f.u.* Thus, the martensitic transformation has great possibility to occur in Pt$_2$MnTi. In addition, if the magnetic moment difference between the tetragonal structure with c/a<1 and the tetragonal structure with c/a>1 is large, then the magnetic field driven strain might also be realized.

3.2 Electronic structure

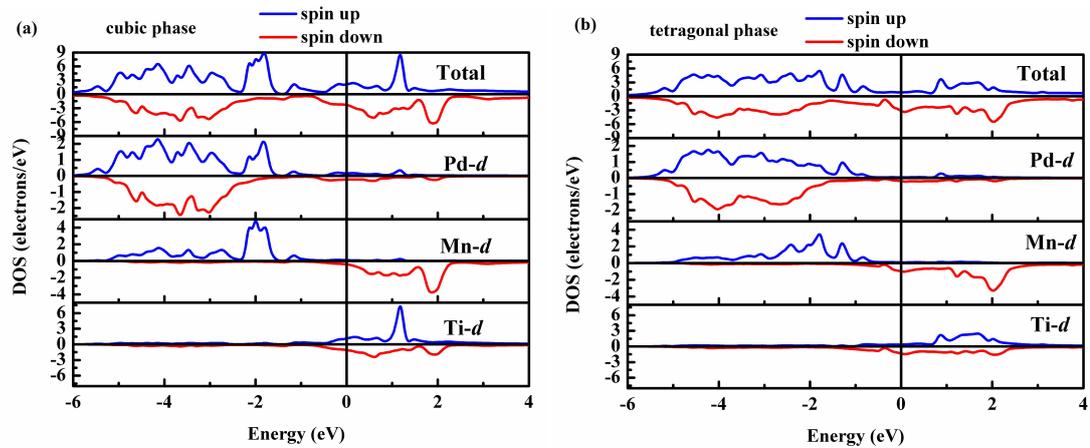

Fig. 4 (a) DOS of cubic phase of Pd$_2$MnTi; (b) DOS of tetragonal phase of Pd$_2$MnTi

In order to explain the reasons for the phase transition from the perspective of electronic structure, the total and partial density of states of both cubic and tetragonal phases of Pd$_2$MnTi have also been drawn in Fig. 4. We have already known that the *p-d* hybridization has a great impact on the stability of the cubic phase of Heusler compounds. From the DOS images, we can find that the *d-d* hybridization also has a very significant impact on the stability of the cubic phase of Heusler compounds. Thus, Ti plays the same role as the main group elements in this system. The detailed analysis is as follows: For the cubic structure, in the spin-up channel, the density of states in the vicinity of the Fermi level is considerable, which brings instability to the cubic phase. And it can be found that these electronic states mainly comes from the Ti-*d* electrons. In addition, there is a obvious peak above the Fermi level at 1.18eV, which is the hybridization peak of Pd-*d* and Ti-*d* electrons. While in the tetragonal phase, in the spin-up channel, the density of state in the vicinity of the Fermi level is greatly reduced and becomes very flat, and the peak which is corresponding to the peak located at 1.18 eV in the cubic phase moves down to 0.86 eV and becomes very weak. In the spin-down channel, the overall density of the electronic states in the vicinity of Fermi level in the tetragonal phase is weakened compared with the cubic phase. However, the small packets near the Fermi level in the spin down channel might also bring instability to the tetragonal phase, so it might imply the possibility to distort from the tetragonal structure to the modulation structure.

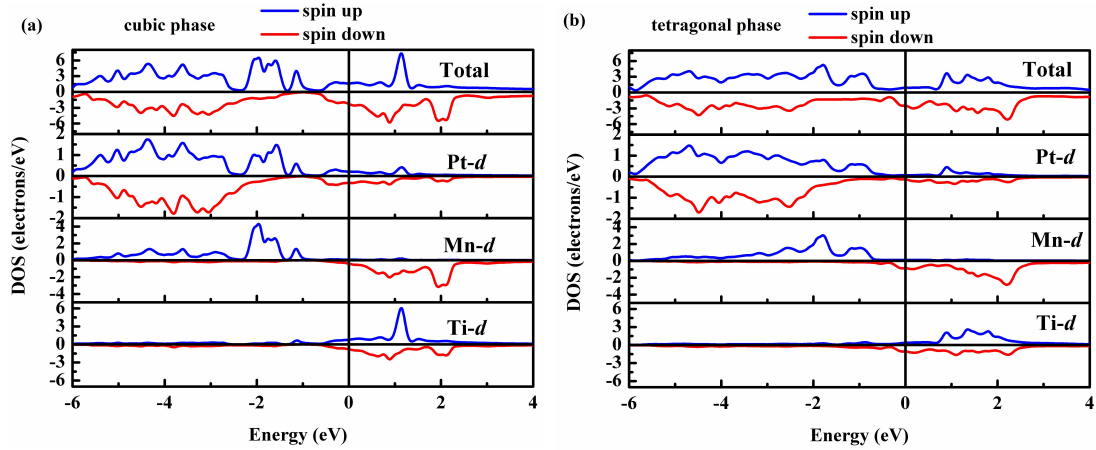

Fig. 5 (a) DOS of cubic phase of $Pt_2MnTi$; (b) DOS of tetragonal phase of $Pt_2MnTi$

The analysis of the electronic structure of $Pt_2MnTi$ is similar to that of $Pd_2MnTi$. The total and partial density of states of both cubic and tetragonal phases of $Pt_2MnTi$ have been drawn in Fig. 5. For the cubic structure, in the spin-up channel, the density of states in the vicinity of the Fermi level is also considerable, which brings instability to the cubic phase. And it can be found that these electronic states mainly comes from the Pt-$d$ and Ti-$d$ electrons. In addition, there is a obvious peak above the Fermi level at 1.13eV, which is the hybridization peak of Pt-$d$ and Ti-$d$ electrons. While in the tetragonal phase, in the spin-up channel, the density of state in the vicinity of the Fermi level is greatly reduced and becomes very flat, and the peak which is corresponding to the peak located at 1.13 eV in the cubic phase moves down to 0.89 eV and becomes very weak. In the spin-down channel, the overall density of the electronic states in the vicinity of Fermi level in the tetragonal phase is also weakened compared with the cubic phase. And the small packets near the Fermi level in the spin down channel might also imply the possibility to distort from the tetragonal structure to the modulation structure.

3.3 Magnetic property

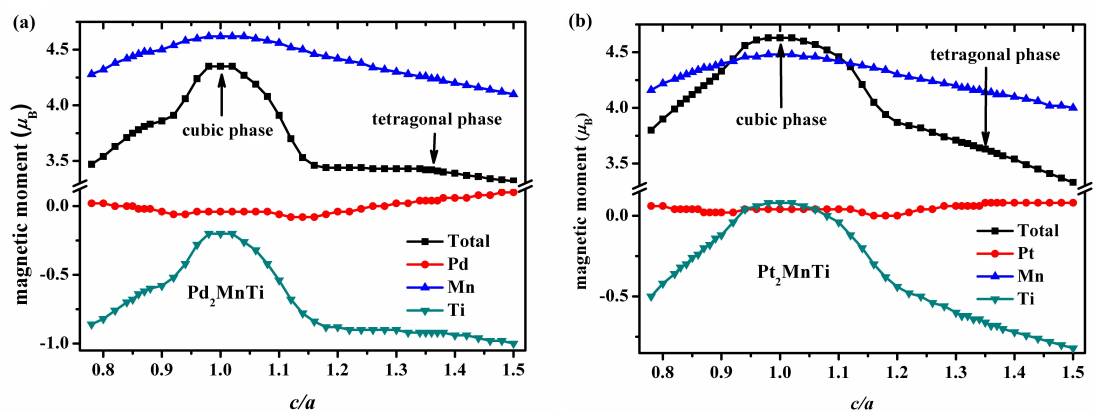

Fig. 6 Variation of the total and atomic magnetic moment of (a) $Pd_2MnTi$ and (b) $Pt_2MnTi$ with $c/a$ ratio.

The magnetic property of $Pd_2MnTi$ and $Pt_2MnTi$ have also been investigated. Table 1 has summarized the total and atomic magnetic moment of the cubic and tetragonal phases of $Pd_2MnTi$. The magnetic moment of Mn change from 4.62 $\mu_B$ in cubic phase to 4.24 $\mu_B$ in the tetragonal phase. And the magnetic moment of Ti is small and negative (-0.20$\mu_B$) in the cubic phase while it become large (-0.92$\mu_B$) in the tetragonal phase. Thus, both the cubic phase and the tetragonal phase are ferrimagnetic. And Fig. 6 (a) has drawn the variation of the total and atomic magnetic moment of $Pd_2MnTi$. It can be found that the magnetic moment of cubic phase is the largest, and the magnetic moment of tetragonal phase decreases with the degree of distortion. The magnetic moment of Pd is very small, which can be almost ignored. And Ti and Mn atoms are coupled antiferromagnetically. The variation of the magnetic moment of Mn is moderate, and the variation of the magnetic moment of Ti is very significant. Thus, the variation of the total magnetic moment of $Pd_2MnTi$ is mainly caused by the variation of the magnetic moment of Ti. This is very different from the variation of the magnetic moment of the main group element in the traditional Heusler compounds. The main group elements in the traditional Heusler compounds only have a very small magnetic moment, and the variation of their magnetic moment with the degree of distortion almost can be ignored.

Table 1 Lattice parameter and magnetic moment of Pd$_2$MnTi, $M_t$, $M_{Mn}$, $M_{Ti}$ and $M_{Pd}$ represent the total magnetic momont, the magnetic moment of Mn, Ti and Pd, respectively.

| Pd$_2$MnTi | $a$(Å) | $c$(Å) | $c/a$ | $M_t(\mu_B)$ | $M_{Mn}(\mu_B)$ | $M_{Ti}(\mu_B)$ | $M_{Pd}(\mu_B)$ |
|---|---|---|---|---|---|---|---|
| austenite | 6.30 | 6.30 | 1.00 | 4.35 | 4.62 | -0.20 | -0.04 |
| martensite | 5.68 | 7.73 | 1.36 | 3.42 | 4.24 | -0.92 | 0.04 |

Table 2 has summarized the total and atomic magnetic moment of cubic and tetragonal phases of Pt$_2$MnTi. It can be found that the magnetic moment of Mn change from 4.48$\mu_B$ in cubic phase to 4.14$\mu_B$ in the tetragonal phase. And the magnetic moment of Ti is very small and positive (0.08$\mu_B$) in the cubic phase while it become considerable and negative (-0.66$\mu_B$) in the tetragonal phase. Thus, the cubic phase is ferromagnetic and the tetragonal phase is ferrimagnetic. And Fig. 6 (b) has also drawn the variation of the total and atomic magnetic moment of Pt$_2$MnTi. It can be found that the magnetic moment of cubic phase is the largest, and the magnetic moment of tetragonal phase decreases with the degree of distortion. The magnetic moment of Pt is very small, which can be almost ignored. In most case, Ti and Mn atoms are coupled antiferromagnetically. The variation of the magnetic moment of Mn is moderate, and the variation of the magnetic moment of Ti is very significant. Thus, the variation of the total magnetic moment of Pt$_2$MnTi is also mainly caused by the variation of the magnetic moment of Ti.

Table 2 Lattice parameter and magnetic moment of Pt$_2$MnTi, $M_t$, $M_{Mn}$, $M_{Ti}$ and $M_{Pt}$ represent the total magnetic momont, the magnetic moment of Mn, Ti and Pt, respectively.

| Pt$_2$MnTi | $a$(Å) | $c$(Å) | $c/a$ | $M_t(\mu_B)$ | $M_{Mn}(\mu_B)$ | $M_{Ti}(\mu_B)$ | $M_{Pt}(\mu_B)$ |
|---|---|---|---|---|---|---|---|
| cubic | 6.32 | 6.32 | 1.00 | 4.63 | 4.48 | 0.08 | 0.04 |
| tetragonal | 5.72 | 7.72 | 1.35 | 3.63 | 4.14 | -0.66 | 0.08 |

4. Conclusions

The martensitic transformation in the new all-*d*-metal Heusler alloys Pd$_2$MnTi and Pt$_2$MnTi have been investigated based on first-principles investigations. The calculated results indicate that the martenstic transformation have great possibility to occur in both Pd$_2$MnTi and Pt$_2$MnTi. The energy differences between the cubic and tetragonal phases are 215.12 meV and 329.45 meV for Pd$_2$MnTi and Pt$_2$MnTi, respectively. The electronic structure analysis indicate that phase stability has been greatly enhanced by the tetragonal distortion. The austenitic phase of Pd$_2$MnTi and Pt$_2$MnTi are ferrimagnetic and ferromagnetic, respectively. And the total magnetic moment of austenitic phase of Pd$_2$MnTi and Pt$_2$MnTi are 4.62 and 4.48 $\mu_B$, respectively.

**Acknowledgement:**

This work is supported by the National Natural Science Foundation of China Grant No. 51301119, the Natural Science Foundation for Young Scientists of Shanxi in Grant No. 2013021010-1, and Scientific and the Technological Innovation Programs of Higher Education Institutions in Shanxi in Grant No. 201802023.

**References:**

[1] Z.Y. Wei, E.K. Liu, J.H. Chen, Y. Li, G.D. Liu, H.Z. Luo, X.K. Xi, H.W. Zhang, W.H. Wang, G.H. Wu, Applied Physics Letters, 107 (2015) 022406.

[2] H.N. Bez, A.K. Pathak, A. Biswas, N. Zarkevich, V. Balema, Y. Mudryk, D.D. Johnson, V.K. Pecharsky, Acta Materialia, 173 (2019) 225.

[3] D.Y. Cong, W.X. Xiong, A. Planes, Y. Ren, L. Manosa, P.Y. Cao, T.H. Nie, X.M. Sun, Z. Yang, X.F. Hong, Y.D. Wang, Physical Review Letters, 122 (2019) 255703.

[4] K.Liu, S.C. Ma, C.C. Ma, X.Q. Han, K. Yu, S. Yang, Z.S. Zhang, Y. Song, X.H. Luo, C.C. Chen, S.U. Rehman, Z.C. Zhong, Journal of Alloys and Compounds, 790 (2019) 78.

[5] Z.Y. Wei, E.K. Liu, Y. Li, X.L. Han, Z.W. Du, H.Z. Luo, G.D. Liu, X.K. Xi, H.W. Zhang, W.H. Wang, G.H. Wu, Applied Physics Letters, 109 (2016) 071904.

[6] S.J. Clark, M.D. Segall, C.J. Pickard, P.J. Hasnip, M.I.J. Probert, K. Refson, M.C. Payne, Z. Krist. 220 (2005) 567.

[7] J.P. Perdew, J.A. Chevary, S.H. Vosko, K.A. Jackson, M.R. Pederson, D.J. Singh, C. Fiolhais, Phys. Rev. B 46 (1992) 6671.

[8] D. Vanderbilt, Phys. Rev. B 41 (1990) 7892.

[9] S.Ö. Kart, M. Uludoğan, I. Karaman, T. Çağın, Phys. Status Solidi A 205 (2008) 1026.